\documentclass[onecolumn,showpacs,preprintnumbers,amsmath,amssymb]{revtex4}
\usepackage{graphics,psfrag}
\usepackage{rotating}
\usepackage[usenames]{color}
\usepackage{amsmath}
\usepackage{bbm}
\usepackage{ulem} %% for strike-through
\newcommand{\be}{\begin{eqnarray}}
\newcommand{\ee}{\end{eqnarray}}

\begin{document}
\title{Chiral dynamics in $\boldsymbol{\gamma p \to \pi^0 \eta p}$ and
related reactions}
\author{M. D\"oring$^{1}$, E. Oset$^{1}$ and D. Strottman$^{1,2}$\\
{\small $^1$ Departamento de F\'{\i}sica Te\'orica and IFIC,
Centro Mixto Universidad de Valencia-CSIC,} \\
{\small Institutos de
Investigaci\'on de Paterna, Aptd. 22085, 46071 Valencia, Spain}\\
{\small $^2$ Theoretical Division, Los Alamos National Laboratory,
Los Alamos, NM
87545}}

\begin{abstract}
Using a chiral unitary approach for meson-baryon scattering in the strangeness zero
sector, where the $N^*(1535)$ and  $\Delta^*(1700)$ resonances are dynamically generated, we study the
reactions $\gamma p \to \pi^0 \eta p$ and $\gamma p \to \pi^0 K^0 \Sigma^+$ at
photon energies at which the final states are produced close to threshold. Among
several reaction mechanisms, we find the most important is the excitation of the
$\Delta^*(1700)$ state 
which subsequently decays into a pseudoscalar meson and the $N^*(1535)$.
Hence, the reaction provides useful
information with which to test current theories of the dynamical generation of the
low-lying $1/2^-$ and $3/2^-$ states. 
Predictions are made
for cross sections and invariant mass distributions which can be compared with
forthcoming experiments at ELSA.
\end{abstract}
\maketitle

\section{Introduction}
The unitary extensions of chiral perturbation theory $U\chi PT$ have brought new
light in the study of the meson-baryon interaction and have shown that some well
known resonances qualify as dynamically generated, or in simpler words, they are
quasibound states of a meson and a baryon, the properties of which are described in
terms of chiral Lagrangians. After early studies in this direction 
explaining the
$\Lambda (1405)$ and the $N^*(1535)$ as dynamically generated resonances
\cite{Kaiser:1995cy,Kaiser:1996js,kaon,Nacher:1999vg,Oller:2000fj}, 
more systematic
studies have shown that there are two octets and one singlet of 
resonances from the
interaction of the octet of pseudoscalar mesons with the octet of 
stable baryons
\cite{Jido:2003cb,Garcia-Recio:2003ks}. The $N^*(1535)$ belongs to 
one of these two
octets and plays an important role in the $\pi N$ interaction with its coupled
channels $ \eta N$, $K \Lambda$ and $K \Sigma$ \cite{Inoue:2001ip}. 
Here, 
we adopt and extend the ideas of Ref. \cite{Nacher:1998mi} for the reaction $\gamma p\to K^+\pi\Sigma$ 
and study the analogous reaction
$\gamma p \to \pi^0 \eta p$ where the $\eta p $ final state can form the $N^*(1535)$
resonance. Besides processes relevant in $\gamma p\to\pi\pi N$ \cite{Nacher:2000eq}, we also include \cite{Doring:2005bx} 
the contribution from the $\Delta^*(1700)$
resonance which qualifies as dynamically generated through the interaction of the $0^-$
meson octet and the $3/2^+$ baryon decuplet as recent studies show
\cite{Kolomeitsev:2003kt,Sarkar:2004jh}. 

At the same time we also study the $\gamma p \to \pi^0 K^0 \Sigma^+$ reaction and
make predictions for its cross section, taking advantage of the fact that it appears
naturally within the coupled channels formalism of the $\gamma p \to \pi^0 \eta p$
reaction and leads to a further test of consistency of the ideas explored here. 
Both reactions are currently being analyzed at ELSA\cite{Metag}. 

\section{Eta pion photoproduction}
Before turning to the two-meson photoproduction, we give a short review of the 
properties of the $N^*(1535)$ in the meson-baryon sector, where this resonance
shows up clearly in the spin isospin $(S=1/2, I=1/2)$ channel.
The model of Ref. \cite{Inoue:2001ip} provides an accurate
data description of elastic and quasielastic $\pi N$ scattering in the $S_{11}$
channel. Within the coupled channel approach in the $SU(3)$ representation of Ref.
\cite{Inoue:2001ip}, not only the $\pi N$ final state is accessible, but also
$K\Sigma$, $K\Lambda$, and $\eta N$ in a natural way.
The corresponding $C_{ij}$ coefficients for the $s$-wave interaction kernel in the charge $Q=+1$ sector are found in Ref. 
\cite{Doring:2005bx}. 
Besides the interaction from the lowest order chiral Lagrangian \cite{ulf, ecker, gasser}, also the $\pi N\to \pi\pi N$
two loop contribution is included.
In Ref. \cite{Doring:2005bx} it is shown that the same model for the resonance 
provides a good data description close to threshold for the one-$\eta$ photoproduction, $\gamma p \to
\eta p$, when including some basic photoproduction mechanisms, which makes us confident 
to use the model of the $N^*(1535)$ for the more complex reaction with two mesons in the final state
in the next section.

\subsection{Production mechanisms for $\boldsymbol{\gamma p\to \pi^0\eta p}$}
\label{sec:production}
\begin{figure}[ht]
\begin{center}
\begin{picture}(300,130)
\put(-45,-5){
\includegraphics[width=12cm]{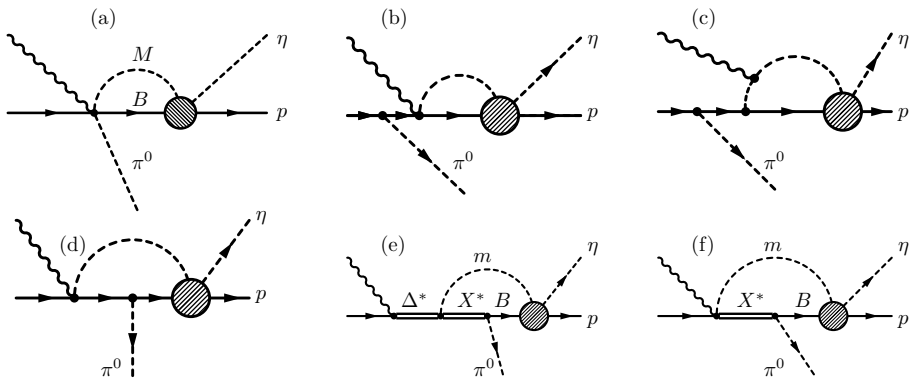}}
\end{picture}
\end{center}
\caption{Photoproduction mechanisms for the $\pi^0\eta p$ final state.}
\label{fig:overview_production}
\end{figure}

In Fig. \ref{fig:overview_production}
the relevant photoproduction mechanisms are shown. Diagrams (c) shows the pole terms that accompanies 
the contact term (b) in order to ensure gauge invariance. Similar terms are included for diagrams (a), (d), 
and (f) but not separately drawn. 
In all diagrams, the gray blob signifies the dynamically generated $N^*(1535)$ that 
provides the final state interaction $MB\to\eta p$ with the intermediate meson-baryon state in the loop 
given by the list of channels from above.
Here, we can only give a qualitative description of the processes and the amplitudes 
which are denoted in detail in Ref. \cite{Doring:2005bx}.

Diagram (a) uses a contact term in a similar way as for the $K^+\Lambda(1405)$ photoproduction in Ref. 
\cite{Nacher:1998mi}, additionally taking 
into account the anomalous magnetic moment of the nucleon with the effective Lagrangian from 
Ref. \cite{Jido:2002yz}. Diagrams (b) and (c) are a straightforward extension of the basic photoproduction 
mechanisms in $\gamma p\to\eta p$ \cite{Doring:2005bx} for the final state with an additional $\pi^0$. It is known that the inclusion
of explicit resonances plays an important role in the two-pion photoproduction 
\cite{Nacher:2000eq}, and therefore we include 
the relevant mechanisms as sub-processes in the calculation:
In diagram (e) we show the contribution from the $\Delta^*(1700)$ decay into $X^*=\Delta(1232)$
and $m=\pi$. In the same way, the $N^*(1520)$ decay from Ref. \cite{Nacher:2000eq} is taken into account.
Fig. (f) shows the corresponding $\Delta$-Kroll-Ruderman term ($X^*=\Delta, m=\pi, B=N$) and also the
$\Sigma^*$-Kroll-Ruderman term ($X^*=\Sigma ^*, m=K, B=\Lambda$).
Additionally, we include the contribution from the dynamically generated $3/2^-$ resonance from Ref. \cite{Sarkar:2004jh}
which has been identified with the $\Delta^*(1700)$. In particular, the decay channels 
$(mX^*B)=(\eta \Delta^+(1232)p),(K^+\Sigma^{*0} (1385)\Lambda),(K^0\Sigma^{*+}\Sigma^+)$ in diagram (e) are possible, whose strengths
are not measured yet in experiment but a genuine prediction of the model  \cite{Sarkar:2004jh}. 
From these couplings, also a tree level diagram originates for the photoproduction: 
$\gamma p\to\Delta^*(1700)\to\eta\Delta^+(1232) [\pi^0 p]$.

\section{Numerical results}
\begin{figure}[ht]
\begin{center}
\begin{picture}(300,150)
\put(-45,-5){
\includegraphics[width=13cm]{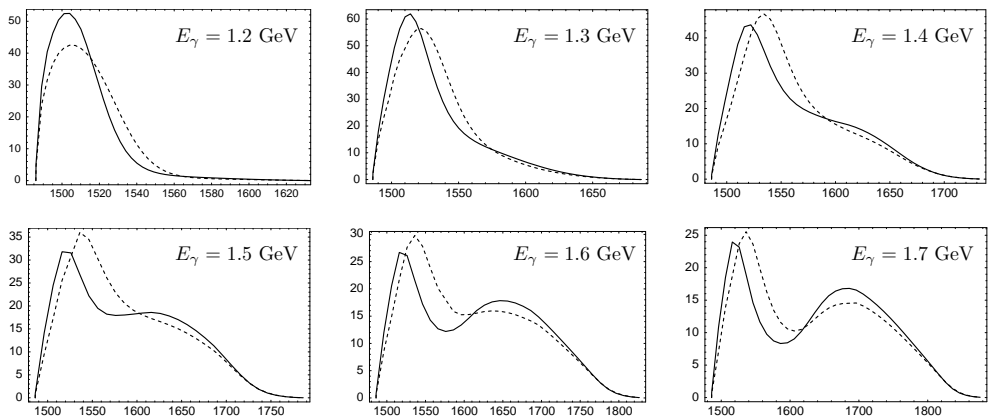}}
\end{picture}
\end{center}
\caption{Invariant mass spectrum $\frac{d\sigma}{dM_I(\eta p)}\;[\mu{\rm b\;GeV}^{-1}]$
as a function of $M_I(\eta p)$ [MeV] for various photon lab energies $E_\gamma$.  Solid and
dashed lines: Full and reduced model for the $N^*(1535)$, respectively.}
\label{fig:bigger_window}
\end{figure}

\begin{figure}[ht]
\begin{center}
\begin{picture}(300,110)
\put(-45,-5){
\includegraphics[width=5.7cm]{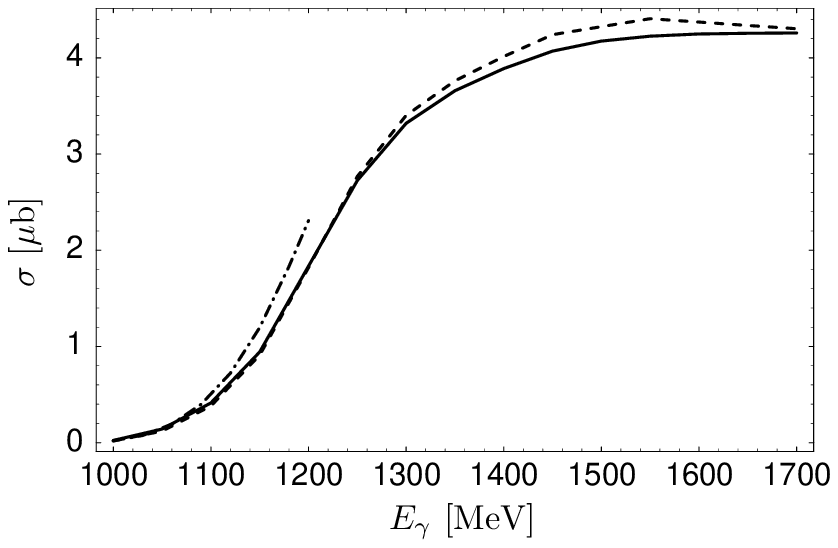}}
\put(140,-5){
\includegraphics[width=6cm]{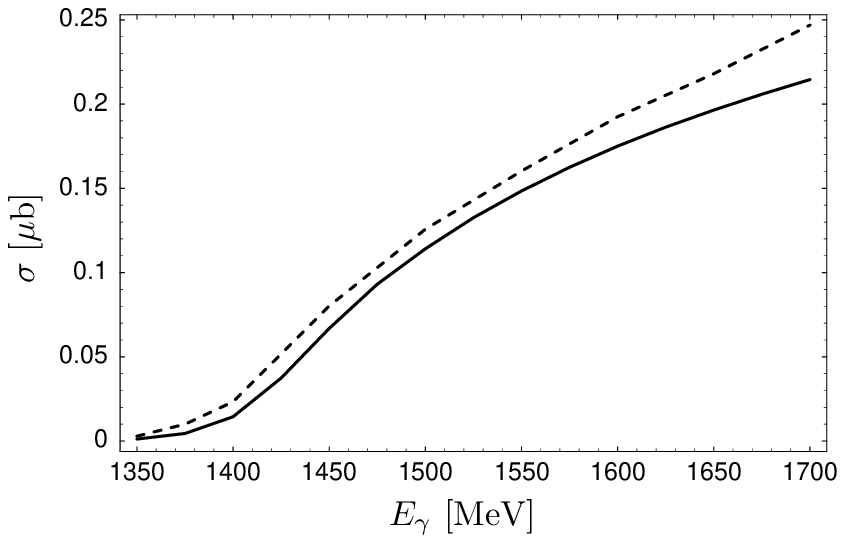}}
\end{picture}
\end{center}
\caption{Left side: Integrated cross section $\sigma$ for the $\gamma p\to \pi^0 \eta p$  reaction. Right side: $\sigma$ for the 
$\gamma p\to\pi^0 K^0 \Sigma^+$ reaction . Solid
line: Full model for the
$N^*(1535)$.  Dashed line: Reduced model. Dashed dotted line (left side): Phenomenological potential
for the $MB\to \eta p$ transition (only available up to $E_\gamma\sim 1.2$ GeV).}
\label{fig:cross_section}
\end{figure}

The invariant mass spectra for $M_I(\eta p)$ and $M_I(\pi^0p)$ and cross section for the $\gamma p\to\pi^0\eta p$ reaction
are predicted which can be directly compared to the forthcoming experiments at the ELSA facility. In Fig. \ref{fig:bigger_window}
the spectrum for $M_{I}(\eta p)$ is shown for several lab photon energies $E_\gamma$, 
taking into account all processes from Fig. \ref{fig:overview_production} plus the tree level process described 
at the end of Sec. \ref{sec:production}. The solid and dashed curves correspond to the full and reduced model
(no $\pi N\to\pi\pi N$ channel, no vector exchange in $t$-channel, see Ref.  \cite{Inoue:2001ip}, \cite{Doring:2005bx})
for the $N^*(1535)$ resonance, and we take the difference as a hint for the theoretical error.
The $N^*(1535)$ is clearly visible in the spectrum, although it interferes with the tree level process at low $E_\gamma$ which 
shifts its position slightly. At higher $E_\gamma$, we observe a second peak which moves with the energy. This is a reflection of the
$\Delta(1232)$ in the tree level process which is on shell around these invariant masses. For a more detailed discussion, see Ref. \cite{Doring:2005bx}.
The $\pi^0p$ invariant mass spectrum is not shown here, but in Ref. \cite{Doring:2005bx}. There, the $\Delta(1232)$ shows up as a clean signal as
expected. The total cross section is plotted in Fig. \ref{fig:cross_section}.
Additionally to the full and reduced model, we show the outcome when using the phenomenological
meson-baryon $\to\eta p$ transition from the PWA from Ref. \cite{Arndt:2003if}. 
From the processes with $N^*(1535)$, diagram (e) from Fig. \ref{fig:overview_production} with $m=\eta$, $X^*=\Delta(1232)$
gives the largest contribution which is due to the strong $\eta p\to\eta p$ transition in the model for 
the $N^*(1535)$. Also, the tree level diagram contributes significantly as we have already seen in Fig. \ref{fig:bigger_window}.
The cross section is relatively independent of the chosen model for the $N^*(1535)$ (solid vs. dashed line), and even the
the phenomenological potential does not change much the result. 

The $\gamma p\to\pi^0K^0\Sigma^+$ reaction is calculated in a similar way as for the  $\pi^0\eta p$ final state by replacing
the $\eta p$ final state of the $N^*(1535)$ by the $K^0\Sigma^+$ one and including a new tree level diagram in analogy to the process
described at the end of Sec. \ref{sec:production}: $\gamma p\to \Delta^{*+}(1700)\to K^0\Sigma^{*+}[\pi^0\Sigma^+]$.
The cross section for $\gamma p\to \pi^0K^0\Sigma^+$ is much smaller than for the $\pi^0\eta p$
final state as Fig. \ref{fig:cross_section} shows. This is --- among other reasons \cite{Doring:2005bx} --- due to the fact that the reaction takes places 
at much higher photon energies where the dynamically generated $N^*(1535)$ is off shell.

\section{Conclusions}
In this paper we have studied the reactions $\gamma p \to \pi^0
\eta p$ and
$\gamma p \to \pi^0 K^0 \Sigma^+$ within a chiral unitary framework  which considers the
interaction of mesons and baryons in coupled channels and  dynamically generates the
$N^*(1535)$. From the various processes, we find dominant the decay of the dynamically
generated $\Delta^*(1700)$
resonance into $\eta \Delta$, followed by the unitarization, or in other words, 
the $\Delta^*(1700)\to \pi^0 N^*(1535)$ decay. A similar term provides also a
tree level process which leads, together with the $N^*(1535)$,  to a characteristic double hump
structure in the $\eta p$ and $\pi^0p$ invariant masses at higher photon energies.
 
A virtue of this approach and a test of the nature of the resonance  as a dynamically
generated object is that one can make predictions about cross sections for the production
of the resonance without introducing the resonance explicitly in the formalism since only
its components in the $(0^-, 1/2^+)$ and $(0^-, 3/2^+)$ meson-baryon base is what matters, together with the coupling of
the photons to these components and their interaction in a coupled channel formalism.
In particular, the reactions studied here probe 
decay channels of these resonances such as $\Delta^*(1700)\to \eta\Delta, \Delta^*(1700)\to K\Sigma^*$ 
or transitions like $\eta p\to N^*(1535)\to\eta p$ which are predicted by the model and not measured yet. 

The measurement of both cross sections is being performed at the ELSA/Bonn Laboratory
and hence the predictions are both interesting and opportune and can  help us gain a
better insight in the nature of some resonances, particularly the 
$N^*(1535)$ and the $\Delta^*(1700)$ in the present case.

\section*{Acknowledgments} This work is partly supported by DGICYT contract number 
BFM2003-00856, and the E.U. EURIDICE network contract no. HPRN-CT-2002-00311. This
research is  part of the EU Integrated Infrastructure Initiative Hadron Physics Project
under  contract number RII3-CT-2004-506078.


\begin{thebibliography}{99}
\bibitem{Kaiser:1995cy}
N.~Kaiser, P.~B.~Siegel and W.~Weise,
%``Chiral dynamics and the S11 (1535) nucleon resonance,''
Phys.\ Lett.\ B {\bf 362} (1995) 23.
%\cite{Kaiser:1996js}
\bibitem{Kaiser:1996js}
N.~Kaiser, T.~Waas and W.~Weise,
%``SU(3) chiral dynamics with coupled channels: Eta and kaon
%photoproduction,''
Nucl.\ Phys.\ A {\bf 612} (1997) 297
%[arXiv:hep-ph/9607459].
%%CITATION = HEP-PH 9607459;%%
%\cite{Oset:1997it}
\bibitem{kaon}
E.~Oset and A.~Ramos,
%``Non perturbative chiral approach to s-wave anti-K N interactions,''
Nucl.\ Phys.\ A {\bf 635} (1998) 99 %[arXiv:nucl-th/9711022].
%%CITATION = NUCL-TH 9711022;%%
%\cite{Nacher:1999vg}
\bibitem{Nacher:1999vg} J.~C.~Nacher, A.~Parreno, E.~Oset, A.~Ramos, A.~Hosaka
and M.~Oka,
%``Chiral unitary approach to the N* N* pi, N* N* eta couplings for the
%N*(1535) resonance,''
Nucl.\ Phys.\ A {\bf 678} (2000) 187
%[arXiv:nucl-th/9906018].
%%CITATION = NUCL-TH 9906018;%%
%\cite{Oller:2000fj}
\bibitem{Oller:2000fj} J.~A.~Oller and U.~G.~Meissner,
%``Chiral dynamics in the presence of bound states: Kaon nucleon interactions
%revisited,''
Phys.\ Lett.\ B {\bf 500} (2001) 263 %[arXiv:hep-ph/0011146].
%%CITATION = HEP-PH 0011146;%%
%\cite{Jido:2003cb}
\bibitem{Jido:2003cb} D.~Jido, J.~A.~Oller, E.~Oset, A.~Ramos and 
U.~G.~Meissner,
%``Chiral dynamics of the two Lambda(1405) states,''
Nucl.\ Phys.\ A {\bf 725}
(2003) 181 %[arXiv:nucl-th/0303062].
%%CITATION = NUCL-TH 0303062;%%
%\cite{Garcia-Recio:2003ks}
\bibitem{Garcia-Recio:2003ks} C.~Garcia-Recio, M.~F.~M.~Lutz and J.~Nieves,
%``Quark mass dependence of s-wave baryon resonances,''
Phys.\ Lett.\ B {\bf 582} (2004) 49 %[arXiv:nucl-th/0305100].
%%CITATION = NUCL-TH 0305100;%%
\bibitem{Inoue:2001ip} T.~Inoue, E.~Oset and M.~J.~Vicente Vacas,
%``Chiral unitary approach to S-wave meson-baryon scattering in the strangeness
%S=0 sector,''
Phys.\ Rev.\ C {\bf 65} (2002) 035204 %[arXiv:hep-ph/0110333].
%%CITATION = HEP-PH 0110333;%%
%\cite{Isgur:1978xj}
\bibitem{Nacher:1998mi} J.~C.~Nacher, E.~Oset, H.~Toki and A.~Ramos,
%``Photoproduction of the Lambda(1405) on the proton and nuclei,''
Phys.\ Lett.\ B {\bf 455} (1999) 55 %[arXiv:nucl-th/9812055].
%%CITATION = NUCL-TH 9812055;%%
\bibitem{Nacher:2000eq}
J.~C.~Nacher, E.~Oset, M.~J.~Vicente and L.~Roca,
%``The role of $\Delta(1700)$ excitation and $\rho$ production in double  pion
%photoproduction,''
Nucl.\ Phys.\ A {\bf 695}, 295 (2001)
%[arXiv:nucl-th/0012065].
%%CITATION = NUCL-TH 0012065;%%
\bibitem{Doring:2005bx}
M.~Doring, E.~Oset and D.~Strottman,
%``Chiral dynamics in the gamma p $\to$ pi^0 eta p and gamma p $\to$ pi^0 K^0
%Sigma^+ reactions,''
arXiv:nucl-th/0510015.
%%CITATION = NUCL-TH 0510015;%%
\bibitem{Kolomeitsev:2003kt}
E.~E.~Kolomeitsev and M.~F.~M.~Lutz,
%``On baryon resonances and chiral symmetry,''
Phys.\ Lett.\ B {\bf 585}, 243 (2004)
%[arXiv:nucl-th/0305101].
%%CITATION = NUCL-TH 0305101;%%
\bibitem{Sarkar:2004jh}
S.~Sarkar, E.~Oset and M.~J.~Vicente Vacas,
%``Baryonic resonances from baryon decuplet - meson octet interaction,''
Nucl.\ Phys.\ A {\bf 750}, 294 (2005)
%[arXiv:nucl-th/0407025].
%%CITATION = NUCL-TH 0407025;%%
\bibitem{Metag} 
V. Metag and M. Nanova, private communication.
\bibitem{gasser} J.~Gasser and H.~Leutwyler, Nucl.\ Phys.\  {\bf B250} (1985)
465, 517, 539.
\bibitem{ulf}U. G. Meissner, Rep. Prog. Phys. {{56}} (1993) 903; V. Bernard, N.
Kaiser and U. G. Meissner, Int. J. Mod. Phys. {{E4}} (1995) 193.
\bibitem{ecker}G. Ecker, Prog. Part. Nucl. Phys. {{35}} (1995) 1.
%%CERES
\bibitem{Jido:2002yz} D.~Jido, A.~Hosaka, J.~C.~Nacher, E.~Oset and A.~Ramos,
%``Magnetic moments of the Lambda(1405) and Lambda(1670) resonances,''
Phys.\ Rev.\ C {\bf 66} (2002) 025203 %[arXiv:hep-ph/0203248].
%%CITATION = HEP-PH 0203248;%%
\bibitem{Arndt:2003if}
R.~A.~Arndt, W.~J.~Briscoe, I.~I.~Strakovsky, R.~L.~Workman and M.~M.~Pavan,
%``Dispersion relation constrained partial wave analysis of pi N elastic  and
%pi N $\to$ eta N scattering data: The baryon spectrum,''
Phys.\ Rev.\ C {\bf 69}, 035213 (2004)
%[arXiv:nucl-th/0311089].
%%CITATION = NUCL-TH 0311089;%%
\end{thebibliography}
\end{document}